\documentclass[prl,twocolumn,preprintnumbers,showpacs]{revtex4}

\usepackage{amsmath} 
\usepackage{amsfonts} 
\usepackage{graphicx}
\usepackage[mathscr]{eucal}

\newcommand{\half}{\tfrac{1}{2}}

\newcommand{\Vol}{\mathop{\mathrm{Vol}}}
\newcommand{\diag}{\mathop{\mathrm{diag}}}

\newcommand{\lag}{\mathcal{L}}
\newcommand{\Mp}{M_{\rm P}}
\newcommand{\Mr}{\hat{M}_{\rm P}}
\newcommand{\MR}{\hat{M}_{\rm R}}
\newcommand{\Lp}{\ell_{\rm P}}
\newcommand{\GN}{G_{\rm N}}
\newcommand{\K}{\mathscr{K}}
\newcommand{\Sn}{\mathbb{S}}
\newcommand{\hb}{\bar{h}}

\newcommand{\beq}{\begin{equation}}
\newcommand{\eeq}{\end{equation}}
\newcommand{\bsub}{\begin{subequations}}
\newcommand{\esub}{\end{subequations}}

\begin{document}

\preprint{gr-qc/0609060}

 \title{Normalization conventions for Newton's constant and the Planck scale in arbitrary spacetime dimension}
\author{Sean P.~Robinson} 
\email{spatrick@mit.edu}
\affiliation{Department of Physics, Massachusetts Institute of Technology,
Cambridge, Massachusetts 02139, USA}
\date{September 16, 2006}

\begin{abstract}
We calculate, in $d$ spacetime dimensions, the relationship between the coefficient $\K^{-2}$ of the Einstein-Hilbert term in the action of general relativity and the coefficient $\GN$ of the force law that results from the Newtonian limit of general relativity. 
The result is $\K^2=2\tfrac{d-2}{d-3}\Vol(\Sn^{d-2})\GN$, where $\Vol(\Sn^n)$ is the volume of the unit $n$-sphere. 
While the $d=4$ case is an elementary calculation in any general relativity text, the arbitrary case presented here is slightly less well known.
 We argue on physical grounds that $\GN$ as defined here is the natural unit of ``square'' surface area (in $\hbar=c=1$ units), such that the Planck length $\Lp\equiv\GN^{1/(d-2)}$, or equivalently the Planck mass $\Mp\equiv 1/\Lp$, is the natural scale of certain quantum gravity phenomena.
 On the other hand, $\K$ sets the coupling strength of linearized metric perturbations to matter and is the relevant factor for comparing the effective strength of gravity in so-called brane world scenarios where gravitation and particle physics occur in spacetimes of differing dimension.
The unitless ratio $\K^2/\GN$ is factorially small for large $d$, potentially disrupting the results of naive dimensional analysis sometimes employed in extra-dimensional particle physics phenomenology. 
However, the ratio of mass scales $(\K^2/\GN)^{1/(2-d)}$ is never more than an order of magnitude from unity for the phenomenologically interesting range of $4\leq d\leq 11$.
The bulk of this effect is due to the volume factor, and is often accounted for by defining a ``reduced Planck scale'' $\Mr\equiv\left[\Vol(\Sn^{d-2})\GN\right]^{1/(2-d)}$, which is the scale for the coupling of the Newtonian potential to matter.
The remaining $2\tfrac{d-2}{d-3}$ factor is a relativistic effect which is rarely accounted for in the literature, but it suggests defining a relativistic reduced Planck scale $\MR\equiv\K^{2/(2-d)}$, which can be lower than $\Mr$ by a factor of up to $1/2$.
\end{abstract}
\maketitle

The constant of universal gravitation $\GN= 6.67\times 10^{-11}$~$\rm{m}^3\rm{kg}^{-1}\rm{s}^{-2}$ \cite{Eidelman:2004wy}, introduced by Newton in 1666 and first measured by Cavendish in 1771, sets the strength of gravitational interactions.
 In Newton's original model, it appears simply as a constant of proportionality in front of a phenomenologically motivated inverse square force law.
 Later developments lead to an understanding of the inverse square law as arising from a more fundamental Gauss law. 
The Gauss law describes the conservation of flux lines as they spread out across the surface of a sphere at ever increasing radii.
 With the benefit of hindsight, it would have been more natural for the denominator of Newton's equation to be not simply $r^2$, but rather the surface area of the sphere $\Vol(\Sn^2)r^2=4\pi r^2$.
 With the insertion of this extra $4\pi$ in the denominator, the overall constant setting the strength of gravitation appears to be $4\pi\GN$. The Gauss law in turn arises as the Newtonian limit --- that is, the limit of both weak fields and nonrelativistic velocities and stresses --- of general relativity. 
In this theory, gravitation is the dynamics of the spacetime metric, and the coupling of metric perturbations to matter is given by $16\pi\GN$.

With the additional consideration of quantum mechanics, one can work in the so-called natural units of particle physics ($\hbar=c=1$) and form the characteristic length scale of quantum gravity, the Planck length $\Lp=\sqrt{\GN}$, and the corresponding Planck mass $\Mp=1/\Lp\approx 1.2\times 10^{19}$~GeV. 
But, the above discussion implies that depending on the theoretical context, one might consider the so-called ``reduced Planck mass'' $\Mr=\Mp/\sqrt{4\pi}\approx 3.4\times 10^{18}$~GeV or an even smaller relativistic reduced Planck mass $\MR=\Mp/\sqrt{16\pi}\approx 1.7 \times 10^{18}$~GeV as the true quantum gravity scale.

The above discussion is unimportant if one is only concerned with the scaling of power laws and dimensional analysis. It remains of little consequence until one considers variations and extensions of Newton's phenomenological model, perhaps motivated by deep theoretical principals.
 One such variation, which is natural to consider following the insight that the physical content of the inverse square law lies in a more fundamental Gauss law, is gravitation in $d-1$ spatial dimensions, rather than the usual three.
 An immediate consequence is that the scaling of the force law with distance changes from $1/r^{2}$ to $1/r^{d-2}$.
 But, in determining the exact form of the force law, coefficients and all, one is faced with a choice of how to insert unitless numerical factors.
For example, should the numerator of Newton's law read $\GN$, $4\pi\GN$, or some $d$-dependent factor times $\GN$?
 Every potential choice is a valid convention as long as it is used consistently.
 However, if one has a theoretical framework in which it makes sense to compare the strength of gravitation in varying dimensions, the prefactor multiplying $\GN$ will enter into the ratios of physical scales.

 For example, many string theory inspired models of particle physics phenomenology have been proposed in recent years in which some or all of the matter fields are confined to a four-dimensional defect, a so-called 3-brane, which is imbedded in a $d$-dimensional bulk spacetime \cite[for example]{Arkani-Hamed:1998rs, Antoniadis:1998ig, Arkani-Hamed:1998nn, Randall:1999ee, Randall:1999vf}. 
 Gravitation is necessarily $d$-dimensional in these scenarios, since it is the dynamics of spacetime itself.
 Some, all, or none of the $d-4$ ``extra'' dimensions can be taken as compactified on a microscopic scale approaching the Planck scale. 
The remainder can be taken as large, sometimes up to millimeter scale, with the details depending on whether the extra dimensions are flat, warped, or otherwise behaving in an interesting way within the given model.
For example, in the simplest case of \cite{Arkani-Hamed:1998rs, Antoniadis:1998ig, Arkani-Hamed:1998nn}, the $d$-dimensional reduced Planck scale $\Mr\phantom{}_{(d)}$ is effectively diluted by the extra-dimensional volume $V_{(d-4)}$ such that the resulting four-dimensional reduced Planck scale is given by $\Mr\phantom{}_{(4)}^2=V_{(d-4)}\Mr\phantom{}_{(d)}^{d-2}$. 
In this case, it has been proposed that while $\Mr\phantom{}_{(4)}\approx 3.4\times 10^{18}~\rm{GeV}$, the fundamental scale $\Mr\phantom{}_{(d)}$ could be made closer to 1 TeV. See \cite{Maartens:2003tw} for a review of brane world phenomenology.

We will examine a convention for $\GN$ in $d$ spacetime dimensions which reflects the physical interpretation of the nonrelativistic gravitational force law as describing the density of diverging, conserved field lines, commensurate with its origin in a Gauss law, which in turn arises as the Newtonian limit of general relativity.
 This convention will be slightly different from others in the literature \cite[for example]{Maartens:2003tw, Zwiebach:2004tj}, but we will use standard $\hbar=c=1$ units.

We start with the simplest generally covariant action for a dynamical spacetime
metric, the Einstein-Hilbert action:
\beq 
S_{\rm{EH}}=\frac{1}{\K^2}\int{d^dx\sqrt{-g}R}
\label{e222.1},
\eeq
where $g$ is the determinant of the metric and $R$ is the Ricci scalar curvature.
The overall coefficient is defined as $1/\K^2$.
Although we will not show it explicitly here, a linearized metric perturbation will have a canonically normalized kinetic term in its Lagrangian when a factor of $1/2\K$ is absorbed into the field definition. 
Accordingly, the units of $1/\K$ are $[\rm{mass}]^{(d-2)/2}$, just like a scalar field. 
After this field redefinition, $\K$ appears in the action as the effective coupling constant between the metric perturbation and any matter in the system. 
It is therefore natural to define a relativistic reduced Planck mass $\MR\equiv 1/\K^{2/(d-2)}$ as the scale of fundamental interactions in quantum gravity.
The quantity more commonly found in the literature, however, is the standard reduced Planck mass $\Mr$, which is defined as the coupling scale of the Newtonian potential to matter.
As we will see below, the Newtonian potential differs from the metric perturbation  by a $d$-dependent factor.
So, $\Mr$  will differ from $\MR$ by a corresponding factor.
We will work with $\K$ for now and return to $\MR$ and $\Mr$ in our conclusions.

  When coupled to matter, the action (\ref{e222.1}) yields the Einstein equation of motion:
\bsub\label{e222.2}\beq
R_{ab}-\half g_{ab}R=\frac{\K^2}{2}T_{ab}
\label{e222.2a},
\eeq
or equivalently 
\beq
R_{ab}=\frac{\K^2}{2}\left(T_{ab}-\frac{1}{d-2} g_{ab}T^c_c\right)
\label{e222.2b},
\eeq\esub
where $R_{ab}$ is the Ricci curvature tensor, $g_{ab}$ is the metric tensor, and indices $a$, $b$, $c$ run over all $d$ spacetime coordinates, $\{0,1,\ldots d-1\}$.
The energy-momentum tensor $T_{ab}$ is defined as
\beq
T_{ab}=2\frac{\delta\lag_{\rm matter}}{\delta g^{ab}}
\label{e222.3}.
\eeq
The factor of 2 in Eq. (\ref{e222.3}), or equivalently the one in the denominator of Eq. (\ref{e222.2a}), is not an arbitrary convention. 
This factor ensures that the energy-momentum tensor so defined --- that is, by variation with respect to the metric, of which there is usually one power in a kinetic matter Lagrangian --- actually corresponds the canonically normalized energy-momentum tensor that is derived as the Noether current of translations---that is, by variations with respect to the matter fields, of which there are usually two powers in a kinetic matter Lagrangian.

To connect to the Newtonian limit, we first need to consider the Einstein equations in the weak field limit, such that we can reliably treat gravitation as a linear perturbation $h_{ab}$ of the metric $g_{ab}$ about a Minkowski background $\eta_{ab}$: $g_{ab}=\eta_{ab}+h_{ab}$, with Minkowski metric $\eta_{ab}=\diag[-1,1\ldots 1]$. 
The tensor $h_{ab}$ has a large gauge symmetry corresponding to general coordinate transformation invariance. 
We choose the well known gauge \beq
\partial_ah^a_b=\half\partial_bh^a_a. \label{gauge}
\eeq
Incidentally, the usefulness of this gauge depends on the factor of $\half$ on the right hand side of Eq. (\ref{gauge}). 
This can be traced back to the fact that the metric is a rank-2 tensor and that its Lie derivative is therefore a sum of two terms, which does not depend on dimension. 
Making the convenient definition $\hb_{ab}\equiv h_{ab}-\half\eta_{ab}h^c_c$ ($h_{ab}=\hb_{ab}-\tfrac{1}{d-2}\eta_{ab}\hb^c_c$), Eq. (\ref{gauge}) reads $\partial_a\hb^{ab}=0$. 
In four dimensions, $\hb_{ab}$ is sometimes called the ``trace reversed'' metric perturbation because $h^c_c=-\hb^c_c$ in that case. 
The trace reverse property does not hold for general $d$, but $\hb_{ab}$ as we have defined it is still a useful quantity.
 Expanding Eq. (\ref{e222.2}) to linear order in $h_{ab}$ and applying the gauge choice, we find:
\bsub\label{e0}\beq
\partial^2\hb_{ab}=-\K^2T_{ab}
\label{e0a},
\eeq
or equivalently 
\beq
\partial^2 h_{ab}=-\K^2\left(T_{ab}-\frac{1}{d-2} \eta_{ab}T^c_c\right)
\label{e0b},
\eeq\esub
where $\partial^2=\eta^{ab}\partial_a\partial_b$ is the d'Alembertian operator.

To reach the Newtonian limit, we go beyond the weak-field limit by further taking the nonrelativistic limit. 
In this limit, $\rho\equiv T_{00}\gg T_{0i}\gg T_{ij}$. 
(Note, indices $i$ and $j$ run over the $d-1$ spatial indices.) 
That is, pressures and fluxes are negligible sources of gravitation in comparison to energy density. 
We also take the nonrelativistic limit to imply that evolution in time is slow in comparison to the spatial extent of matter distributions, such that time derivatives can be ignored with respect to spatial gradients.
Demanding regular solutions of Eq. (\ref{e0a}) then gives 
\bsub\label{ea00}
\beq\hb_{0i}\approx\hb_{ij}\approx 0 \label{ea0}
\eeq
and
\beq
\nabla^2\hb_{00}=-\K^2\rho, \label{ea}
\eeq
\esub
where $\nabla^2$ is the Laplacian in $(d-1)$-dimensional Euclidean space. 
Since $\hb_{00}$ is the only nonzero component of $\hb_{ab}$, we also find $\hb^c_c=-\hb_{00}$. 
Thus, 
\bsub\label{ex}
\begin{align}
h_{00}=&\tfrac{d-3}{d-2}\hb_{00},\label{exa} \\ 
h_{0i}=&0,\label{exb} \\ 
h_{ij}=&\tfrac{1}{d-2}\delta_{ij}\hb_{00} \label{exc}, \end{align}\esub
where $\delta_{ij}$ is the $(d-1)$-dimensional Kronecker $\delta$.

 Having considered the dynamics of gravity in the Newtonian limit, we must now examine the motion of test masses under the influence of gravity, but no other forces, in the same limit.
The trajectory of such a test mass is given by a geodesic curve. 
This is a curve $x^a(\tau)$ --- where $x^a$ are $d$-dimensional spacetime coordinates and $\tau$ is proper time along the curve --- which parallel transports its own tangent vector $u^a=dx^a/d\tau$:
\beq
\frac{du^a}{d\tau}+\Gamma^a_{bc}u^bu^c \label{e1}=0,
\eeq
where $\Gamma^a_{bc}$ is the Christoffel connection.
In the nonrelativistic limit, $\tau\approx x^0\equiv t$, $u^0\approx 1$, and $u^i\approx 0$. 
So, the spatial part of the geodesic equation becomes
\beq
\frac{d^2x^i}{dt^2}=-\Gamma^i_{00} = a^i\label{e2},
\eeq
where $a^i$ is the test particle's coordinate acceleration.
Applying the Newtonian limit to the Christoffel connection in Eq. (\ref{e2}), we get
\beq
a^i=\frac{1}{2}\partial^ih_{00} \label{e3}.
\eeq
Thus, the acceleration $a^i$ of a particle under the influence of an approximately Newtonian gravitational field is given by minus the gradient of a potential 
\beq\varphi\equiv-\half h_{00}\label {e3.5},\eeq
such that
\beq
a^i=-\partial^i\varphi \label{e4}.
\eeq

As an aside, we note a second derivation of Eq. (\ref{e3}). 
In Newtonian kinematics, the relative separation $y^i$ of two nearby particles in an acceleration field $a^i$ is given by $d^2y^i/dt^2=y^j\partial_ja^i$, the tidal acceleration. 
The corresponding quantity for two nearby particles in geodesic motion on a curved spacetime is given the equation of geodesic deviation, $d^2y^a/d\tau^2=-y^c{R_{bcd}}^au^bu^d$, where $u^a$ is the tangent vector along the geodesics and ${R_{bcd}}^a$ is the full Riemann curvature tensor.
 In the same Newtonian limit used above, we find $d^2y^i/dt^2\approx-y^j{R_{0j0}}^i\approx \half y^j\partial_j\partial^i h_{00}$. 
This again suggests, as in Eq. (\ref{e3}), that $a^i=\half\partial^ih_{00}$. 
Of course, the two arguments --- one in terms of geodesic motion and connections and the other in terms of geodesic deviation and curvature --- are only superficially different. 
More importantly, however, both arguments are purely kinematical. 
At no point do they invoke any dynamical content of the Einstein equations or any other theory of spacetime dynamics. 

\begin{figure}
\includegraphics[width=3in]{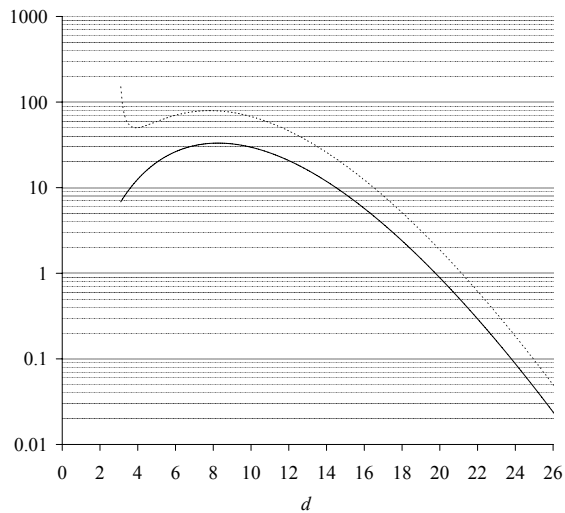}
\caption{\label{fig1} Two $d$-dependent quantities are plotted in the range $3<d<26$. The solid line is $\Vol(\Sn^{d-2})$, and the dotted line is $\K^2/\GN$ as given in Eq. (\ref{e222.6}). Axes are unitless.}
\end{figure}

Combining Eqs. (\ref{ea}, \ref{exa}, and \ref{e3.5})  we get 
\beq 
\nabla^2\varphi=\frac{\K^2(d-3)}{2(d-2)}\rho
\label{e222.4}.
\eeq
Integrating Eq. (\ref{e222.4}) for a point mass $M$ stationary at the origin --- that is, $\rho=M\delta^{(d-1)}(x^i)$, where $\delta^{(n)}(x)$ is the $n$-dimensional Dirac $\delta$ --- gives an acceleration field for a test mass of 
\beq
|a^i|=\frac{M}{\Vol{(\Sn^{d-2})}r^{d-2}}\frac{\K^2(d-3)}{2(d-2)}
\label{e222.5}
\eeq 
in the inward radial direction.
Of course, $\Vol(\Sn^{d-2})=\frac{2\pi^{(d-1)/2}}{\Gamma((d-1)/2)}$ (see Fig. \ref{fig1}).

Since the physical context of the $1/r^{d-2}$ law is that there is a total flux generated at some small distance which is conserved as it spreads out over the area of a sphere at a larger distance $r$, we take Eq. (\ref{e222.5}) to represent the ratio of the areas of the spheres at these different distances. 
So, the factor $\frac{\K^2(d-3)}{2(d-2)}$ should be thought of as the area of a sphere of some special unit radius, $\frac{\K^2(d-3)}{2(d-2)}\equiv\Vol{(\Sn^{d-2})}\Lp^{d-2}$. 
We then define $\GN\equiv\Lp^{d-2}\equiv\Mp^{2-d}$ as the basic unit for counting square areas. 
Thus,
\beq
\K^2=2\frac{d-2}{d-3}\Vol{(\Sn^{d-2})}\GN
\label{e222.6}
\eeq
(again, see Fig. \ref{fig1}).
With this choice for the $d$ and $\GN$ dependence of $\K$, the Newtonian acceleration field reads $|a^i|=M\frac{\Vol\left(\Sn^{d-2}_{\Lp}\right)}{\Vol\left(\Sn^{d-2}_r\right)}$, where $\Vol\left(\Sn^n_r\right)$ is the volume of an $\Sn^n$ of radius $r$. 

\begin{figure}
\includegraphics[width=3in]{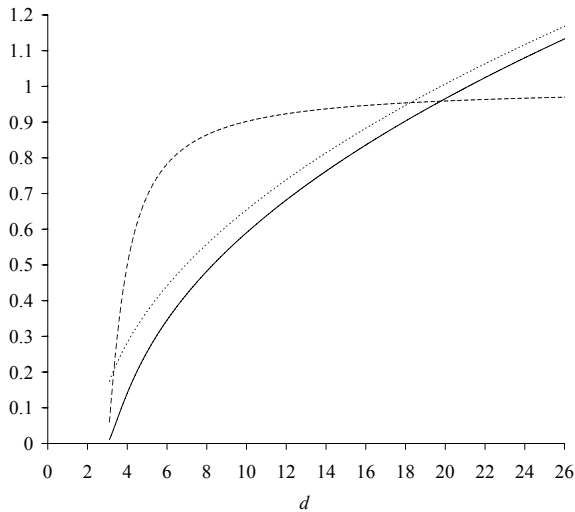}
\caption{\label{fig2} Ratios of assorted Planck scales are plotted in the range $3<d<26$. The solid line is $\MR/\Mp$, the dotted line is $\Mr/\Mp$, and the dashed line is $\MR/\Mr$. Axes are unitless.}
\end{figure}

Comparing Eq. (\ref{e222.6}) to the definitions of the Planck mass and relativistic reduced Planck mass, we find
\beq
\MR=\left(\frac{d-3}{2(d-2)}\frac{1}{\Vol(\Sn^{d-2})}\right)^{1/(d-2)}\Mp.
\label{e10}
\eeq
The ratio $\MR/\Mp$ is a monotonically increasing function of $d$, scaling as approximately $\sqrt{d}$ at very large $d$ (see Fig. \ref{fig2}).
The ratio's value is less than 1 for $d<22$, but only moderately so; it's smallest nonzero value is only $1/(4\sqrt{\pi})\approx 0.14$ at $d=4$. 
So, for the typical regime of interest for phenomenology, $4\leq d\leq 11$, whether one uses $\Mp$ or $\MR$ is not of great numerical significance. 
Some authors, such as \cite{Arkani-Hamed:1998nn}, omit the factor $\tfrac{d-3}{2(d-2)}$ from their definition of $\MR$, leaving what is often denoted $\Mr$. 
These authors are using the coupling of the Newtonian potential, rather than the relativistic metric perturbation, to define the Planck scale.
This distinction is stated explicitly in \cite{Arkani-Hamed:1998nn}, but many authors are not so careful.
The result is that $\MR$ is even more suppressed than $\Mr$ relative to $\Mp$. 
This effect is greatest in lower dimensions, taking its strongest value of $\half$ in $d=4$. 
The value asymptotes to 1 for large $d$ (again, see Fig. \ref{fig2}).

We have calculated the relationship between the coefficients which set the interaction strengths of classical general relativity and of it's Newtonian (weak field, low velocity) limit. The result is Eq. (\ref{e222.6}):
\beq
\K^2=2\frac{d-2}{d-3}\Vol{(\Sn^{d-2})}\GN
\label{e11}. \nonumber
\eeq
We argued on physical grounds that the Planck length and mass should be defined as $\Lp=\GN^{1/(d-2)}$ and $\Mp=\GN^{1/(2-d)}$ respectively, in agreement with typical definitions in the literature. 
Eq. (\ref{e222.6}) defines a natural relativistic reduced Planck scale via $\K^2=\MR^{2-d}$, which is physically relevant because $\K$ is the coupling strength of metric perturbations to matter. 
We noted that some authors define a reduced Planck scale based on the coupling strength of the Newtonian potential to matter, which can differ from $\MR$ by a factor of up to $\half$.
While each of $\Mp$, $\Mr$, and $\MR$ exhibit regimes where they are computationally convenient and physically relevant, $\MR$ is the one which appears most directly in the action of general relativity.
In that sense it can be regarded as more fundamental than the others.
 

\end{document}